\documentclass[preprint,english,showpacs,12pt,floatfix]{revtex4-1}
\usepackage{babel,amsmath,amssymb,dcolumn}
\usepackage{hyperref}
\usepackage{amsfonts}
\usepackage{amssymb}
\usepackage{graphicx}
\usepackage{graphics}
\usepackage{latexsym}
\usepackage{dcolumn}
\usepackage{epsfig}
\usepackage{subfigure}
\usepackage{array}
\usepackage{multirow}
\usepackage[retainorgcmds]{IEEEtrantools}
\allowdisplaybreaks[1]
\begin{document}

\newcommand{\vp}{\varphi}
\newcommand{\nn}{\nonumber\\}
\newcommand{\beq}{\begin{equation}}
\newcommand{\eeq}{\end{equation}}
\newcommand{\bed}{\begin{displaymath}}
\newcommand{\eed}{\end{displaymath}}
\def\bea{\begin{eqnarray}}
\def\eea{\end{eqnarray}}
\newcommand{\veps}{\varepsilon}
\newcommand{\nablasl}{{\slash \negthinspace \negthinspace \negthinspace \negthinspace  \nabla}}
\newcommand{\om}{\omega}

\newcommand{\Dsl}{{\slash \negthinspace \negthinspace \negthinspace \negthinspace  D}}
\newcommand{\tDsl}{{\tilde \Dsl}}
\newcommand{\tnablasl}{{\tilde \nablasl}}
\title{Quantum stress tensor of massive vector field in the space-time of a pointlike global monopole}

\author{Owen Pavel Fern\'{a}ndez Piedra$^{1, 2}$ \\
\textit{$^1$ Departamento de F\'isica, Divisi\'on de Ciencias e Ingenier\'ias, Universidad de Guanajuato, Campus Le\'on, Loma del Bosque N0. 103, Col. Lomas del Campestre, CP 37150, Le\'on, Guanajuato, M\'exico.}\\
\textit{$^2$Grupo de Estudios Avanzados, Universidad de Cienfuegos, Carretera a Rodas, Cuatro Caminos, s/n. Cienfuegos, Cuba.}}
\email{opavelfp2006@gmail.com}

\begin{abstract}
The components of the renormalized quantum stress tensor for a massive vector field in the spacetime of a pointlike global monopole are determined analytically in the Schwinger-DeWitt approximation. The general results are employed to investigate the pointwise energy conditions for the quantized matter field, and it is shown that they are all violated by the massive vector field.
\end{abstract}
\date{\today}
\maketitle

\section{Introduction}
Phase transitions in the early universe due to the global symmetry breaking have had tangible cosmological consequences, highlighting an important relationship between particle physics and cosmology. One of these cosmological consequences has been the formation, after Planck's time, of macsoscopic topological defects, in the form of domain walls, monopoles and cosmic strings \cite{kibble,vilenkin}.

An interesting particular case is that related to global monopoles, first discovered by Sokolov and Starobinski \cite{sokolov}. The most simple model of this system is due to Barriola and Vilenkin \cite{barriola-vilenkin}, who consider this heavy topological defect predicted by Grand-Unification theories, as a consequence of a phase transition of a self-coupling scalar field triplet, due to the spontaneous breaking of original global O(3) symmetry to U (1). The scalar fields plays the role of an order parameter which acquires a non zero value outside the monopole's core.

In the model of Barriola and Vilenkin, global monopoles have, far from the monopole's center, geometry similar to that of a black hole, but with a solid angle deficit. Neglecting the mass term of the monopole, Barriola and Vilenkin obtain the metric of a pointlike global monopole, given by \cite{barriola-vilenkin}:
\begin{equation}
ds^2 = -\alpha^2 dt^2 + dr^2/\alpha^2 + r^2(d\theta^2 +
\sin^2\theta d\varphi^2)\ ,
\label{metric}
\end{equation}
where we define the parameter $\alpha$ according to the expression $\alpha^2 = 1 - 8\pi \eta^2$, with $\eta$ of order $10^{16}Gev$ for a typical Grand Unified theory.

Re-scaling the $t$ and $r$ variables in (\ref{metric}) as $\tau=\alpha t$ and $\rho=\frac{r}{\alpha}$ we obtain
\begin{equation}
ds^2 = -d\tau^2 + d\rho^2 + \alpha^{2}\rho^2(d\theta^2 +
\sin^2\theta d\varphi^2)\ ,
\label{metric2}
\end{equation}
which shows that, as $\alpha< 1$, (\ref{metric2}) represent an spacetime with a solid angle deficit, defined as the difference between the solid angle in the flat spacetime $4\pi$ and the solid angle in the global monopole spacetime $4\pi\alpha^2$.

An interesting problem to be considered is the influence of the nontrivial topology of the pointlike global monopole spacetime upon the vacuum expectation values of physical observables in a quantum field theoretical model of this system. In the absence of a full quantum theory of gravity, we can consider Quantum Field theory in Curved spacetime as a useful alternative to address the above mentioned problem. Quantum Field theory in Curved spacetime considers matter fields that obeys to the laws of quantum theory but propagating in a gravitational background considered as classical \cite{birrel}. In this theory, two important physical quantities offer all the information concerning matter fields. This quantities are the renormalized stress energy tensor of the quantum field \(\langle T_{\mu}^{\nu}\rangle_{ren}\), and the vacuum expectation value of the square of the field $\left<\phi^{2}\right>$, also called vacuum polarization. This two quantities can be derived, in principle, using a variety of mathematical techniques, including analytical, semianalytical and numerical ones \cite{birrel}-\cite{Folacci}.

The quantization of fields around global monopole have been addressed in references \cite{hiscock,mazzitely-lousto,bezerra1}, where the massless scalar field is investigated, and in \cite{bezerra2,bezerra3}, where the quantum stress energy tensor for a massless and massive spinor fields are determined.

In three previous papers, we begin a program to study the quantization of massive fields of arbitrary spin in the pointlike global monopole spacetime, using the Schwinger-DeWitt proper time formalism. We first consider a massive scalar field with arbitrary coupling to the gravitational field of a pointlike global monopole, finding analytical expressions for the vacuum polarization $\left<\phi^{2}\right>$  and the renormalized stress tensor \cite{owenmonopole1,owenmonopole2}. The quantum stress tensor for the massive spinor field was considered in our paper \cite{owenmonopole3}. In this report, we will present the results of our study concerning the quantization of massive vector field, giving analytical formulae for the renormalized stress energy tensor, for a general spacetime, and for the specific case of a pointlike global monopole background.

We will use the Schwinger-DeWitt proper time technique, which is adequate to treat the case of massive fields, whose equation of motion come given in terms of a "minimal" second order differential operators of the general form:
\begin{equation}\label{}
     \hat{K}^{\mu}_{\nu}\left(\nabla\right)=\delta_{\nu}^{\mu}\Box-m^{2}\delta_{\nu}^{\mu}+Q^{\mu}_{\nu} \label{minimal}
\end{equation}
where \(Q^{\mu}_{\nu}(x)\) is some arbitrary matrix playing the role of the potential, and $m$ is the mass of the field.

In this formalism, we can obtain an expansion of the renormalized effective action for the massive quantum field of mass $m$, in terms of the square of inverse mass $\frac{1}{m^{2}}$. In four dimensions, the first three terms in the expansion renormalize the bare gravitational and cosmological constant, and adds some higher order terms to the Einstein gravitational action. The next order term, is the leading term of the one-loop effective action $W_{ren}$ for the matter field \cite{frolov,DeWitt,avramidi,barvinsky,matyjasek,matyjasek1,owencabo1,owencabo2}, and defines the Schwinger-DeWitt approximation, that will be use in this paper, expected to be valid expect whenever the Compton's wavelenght of the field is less than the characteristic radius of curvature of the spacetime. Higher order terms with respect to the leading one, can in principle be interest to be considered to get better results, but in view of the extremely complicate calculations involved, we not include them in this paper, leaving open this problem to be considered in future works.

By functional differentiation of the one-loop effective action, we can obtain the desired quantum stress tensor using the standard
formula
\begin{equation}\label{}
    \langle T_{\mu\nu}\rangle_{ren}=\frac{2}{\sqrt{\ -g}}\frac{\delta W_{ren}}{\delta g^{\mu\nu}}
\label{setdef}
\end{equation}

In the following consider a single massive neutral vector field with mass $\mu$ in the gravitational background of a pointlike global monopole in four dimensions. The action for the system is \cite{deWitt2}:
\begin{equation}
    S=-\int d^{4}x\sqrt{-g}\left(\frac{1}{4}F_{\mu\nu}F^{\mu\nu}+\frac{1}{2}m^{2}A_{\mu}A^{\mu}\right)\label{action}
\end{equation}
The equation of motion for the field have the form
\begin{equation}
    \hat{D}^{\mu}_{\nu}\left(\nabla\right)A_{\mu}=0 \label{fieldeqn1}
\end{equation}
where the second order operator \(\hat{D}^{\mu}_{\nu}\left(\nabla\right)\) is given by
\begin{equation}\label{}
    \hat{D}^{\mu}_{\nu}\left(\nabla\right)=\delta_{\nu}^{\mu}\Box-\nabla_{\nu}\nabla^{\mu}-R_{\nu}^{\mu}-m^{2}\delta_{\nu}^{\mu}\label{nonminimal}
\end{equation}
being  $\Box\,=\,g^{\mu\nu}\nabla_{\mu}\nabla_{\nu}$ the covariant D'Alembert operator, and \(\nabla_{\mu}\) the covariant derivative.

The one-loop effective action, in terms of the operator (\ref{nonminimal}) is given by:
\begin{equation}\label{}
    \Gamma_{(1)}=\frac{i}{2}\ln\left(\mathfrak{Det}\hat{D}\right)
\end{equation}
where \(\mathfrak{Det}\hat{F}=\exp(\mathbb{T}\mathrm{r}\ln\hat{F})\)
defines the functional Berezin superdeterminant \cite{avramidi} of the
operator \(\hat{F}\), and \(\mathbb{T}\mathrm{r}
\hat{F}=\left(-1\right)^{i}F^{i}_{i}=\int
d^{4}x\left(-1\right)^{A}{F}^{A}_{A}(x)\) defines the functional
supertrace \cite{avramidi}.

Is it easy to see that the presence in (\ref{nonminimal}) of the nondiagonal term in is an obstacle to applying the Schwinger-DeWitt technique, valid for operator of the form (\ref{minimal}). But taking into account the identity
$\hat{D}^{\mu}_{\nu}\left(\nabla\right)\left(m^{2}\delta_{\nu}^{\mu}-\nabla_{\nu}\nabla^{\mu}\right)=m^{2}\left(\delta_{\nu}^{\mu}\Box-R_{\nu}^{\mu}-m^{2}\delta_{\nu}^{\mu}\right)$, the one-loop effective action for the nonminimal operator (\ref{nonminimal}) can be written, omitting an inessential constant, as
\begin{equation}\label{}
   \frac{
   i}{2}\mathbb{T}\mathrm{r}\ln\hat{D}^{\mu}_{\nu}\left(\nabla\right)=\frac{
   i}{2}\mathbb{T}\mathrm{r}\left(\delta_{\nu}^{\mu}\Box-R_{\nu}^{\mu}-m^{2}\delta_{\nu}^{\mu}\right)-\frac{
   i}{2}\mathbb{T}\mathrm{r}\left(m^{2}\delta_{\nu}^{\mu}-\nabla_{\nu}\nabla^{\mu}\right)\label{split}
\end{equation}
We can identify the first term in (\ref{split}) as the effective action of a minimal second order operator
\(K^{\mu}_{\nu}\left(\nabla\right)\) with potential \(-R_{\nu}^{\mu}\). The second term can be transformed as
\(\mathbb{T}\mathrm{r}\left[\frac{1}{m^{2}}\nabla^{\mu}\nabla_{\nu}\right]^{n}=\mathbb{T}\mathrm{r}\left[\frac{1}{m^{2}}\nabla^{\mu}\Box^{n-1}\nabla_{\nu}\right]=\mathbb{T}\mathrm{r}\left[\frac{1}{m^{2}}\Box\right]^{n}\)
which gives $\frac{i}{2}\mathbb{T}\mathrm{r}\left(m^{2}\delta_{\nu}^{\mu}-\nabla_{\nu}\nabla^{\mu}\right)=-\frac{i}{2}\mathbb{T}\mathrm{r}\left(\Box-m^{2}\right)$, which is he oposite of the effective action os a minimally coupled massive scalar field. Then, the effective action for the massive vector field is simply equal to the effective action of the minimal second order operator \(K^{\mu}_{\nu}\left(\nabla\right)\) minus the effective action of a massive scalar field minimally coupled to gravity.

Now using the Schwinger-DeWitt representation for the Green's function of the operator (\ref{minimalop}), we can obtain the renormalized one-loop effective action of the quantum field \(\phi\):
\begin{equation}\label{}
         W^{(1)}_ {ren}\,=\,{1\over 32\pi^{2}\,}\,\int d^{4}x \sqrt{-g}\,\sum_{k=3}^{N}\frac{\left(k-3\right)!}{m^{2(k-2)}}\left[a_{k}\right]
\label{efactiongral}
\end{equation}

where $[a_{k}]= \lim_{x'\rightarrow x}\,a_{k}(x,x')$, are the coincident limits of the Hadamard-DeWitt coefficients, that are purely geometric terms whose complexity rapidly increases with \(k\). Expressions for this coefficients up to order $k=4$ are given in reference \cite{avramidi}. 

Due to the many identities satisfied by the Riemann tensor, its contractions and covariant derivatives, the expressions for the Hadamard-DeWitt coefficients, and as a consequence the one-loop effective action (\ref{efactiongral}) and the renormalized stress tensor given by (\ref{setdef})are not unique, but all must be equivalent. In a paper by Decaninis and Folacci they addressed propose a solution to this problem by use of a specific base of scalar constructed from the above tensors \cite{Folacci}. In the following we use this basis for display the expressions for the effective action and the renormalized stress energy tensor of a massive vector field.

Using integration by parts and the elementary properties of the Riemann tensor, we can show that the one-loop effective action of a massive vector field can be written as

\begin{eqnarray} \label{efactionok}
& & W^{(1)}_{\mathrm{ren}}= \frac{1}{192 \pi^2 m^2}\int d^4
x\sqrt{-g} \left[-\frac{27}{280}
  \, R\Box R  +\frac{9}{28}\,  R_{\alpha \beta} \Box R^{\alpha \beta} -\frac{5}{72} \, R^3  + \frac{31}{60}\, RR_{\alpha \beta} R^{\alpha \beta}  \right. \nonumber \\
& & \qquad \left.   - \frac{52}{63} \, R_{\alpha \beta} R^{\alpha}_{\phantom{\alpha} \gamma}R^{\beta \gamma} -\frac{19}{105}
\, R_{\alpha \beta}R_{\gamma \delta}R^{\alpha \gamma \beta \delta}- \frac{1}{10} \, R_{\alpha \beta}R^\alpha_{\phantom{\alpha} \gamma \delta \epsilon} R^{\beta \gamma \delta \epsilon} \right. \nonumber \\
& &  \qquad \left.
       +\frac{61}{140}\, RR_{\alpha \beta \gamma \delta} R^{\alpha \beta \gamma \delta}+ \frac{67}{2520} \, R_{\alpha \beta \gamma \delta}R^{\alpha \beta \sigma \rho}
R^{\gamma \delta}_{\phantom{\gamma \delta} \sigma \rho} + \frac{1}{18}\, R_{\alpha \gamma \beta \delta} R^{\alpha \phantom{\sigma}
\beta}_{\phantom{\alpha} \sigma \phantom{\beta} \rho} R^{\gamma \sigma \delta \rho} \right].
\end{eqnarray}

The renormalized quantum stress energy tensor for the massive vector field in a generic spacetime background can be determined from (\ref{efactionok}) by functional differentiation with respect to the metric tensor:
\begin{eqnarray}\label{SET}
&&  \langle  ~T_{\mu\nu
} ~  \rangle_{\mathrm{ren}}=\frac{2}{ \sqrt{-g}} \frac{\delta
W^{(1)}_{\mathrm{ren}}} {\delta
g_{\mu\nu }}= \frac{1}{96 \pi^2 m^2}\left[\frac{9}{70} \, (\Box R)_{;\mu\nu}
- \frac{7}{60}\, R \Box R_{\mu \nu} + \frac{13}{35}\,R_{;\alpha (\mu} R^{\alpha}_{\phantom{\alpha}\nu)}-\frac{9}{28} \,  \Box \Box R_{\mu \nu} \right. \nonumber\\
&& \qquad \qquad \left.  -\frac{1}{10}\,  R R_{;\mu \nu}-\frac{7}{30}\, (\Box R) R_{\mu \nu} + \frac{337}{210} \, R_{\alpha (\mu}
\Box
R^\alpha_{\phantom{\alpha} \nu)}+ \frac{22}{105} \,R^{\alpha \beta} R_{\alpha \beta;(\mu \nu)}+ \frac{34}{105} \, R^{\alpha \beta} R_{\alpha (\mu ; \nu)\beta}\right.\nonumber \\
&& \qquad \qquad \left.
 -\frac{107}{210} \, R^{\alpha \beta} R_{\mu\nu; \alpha \beta}+\frac{1}{21} \,R^{;\alpha \beta}R_{\alpha \mu \beta \nu}  -\frac{22}{35} \, (\Box R^{\alpha \beta})R_{\alpha \mu \beta \nu}+
 \frac{46}{35} \,R^{\alpha \beta;\gamma}_{\phantom{\alpha \beta;\gamma} (\mu} R_{|\gamma \beta \alpha|
\nu)}   \right. \nonumber \\
&& \qquad \qquad \left. + \frac{116}{105}  \,R^{\alpha \phantom{(\mu}; \beta \gamma}_{\phantom{\alpha } (\mu} R_{|\alpha \beta \gamma|
\nu)} -\frac{1}{42} \,R^{\alpha \beta \gamma \delta} R_{\alpha \beta \gamma \delta ; (\mu \nu)
}-\frac{1}{24}\, R_{;\mu} R_{;\nu} +\frac{83}{210} \,
R_{;\alpha } R^\alpha_{\phantom{\alpha} (\mu;\nu)}  - \frac{41}{84} \,R_{;\alpha }R_{\mu\nu}^{\phantom{\mu\nu}; \alpha} \right. \nonumber \\
&& \qquad \qquad \left.
+\frac{31}{60}\, R^{\alpha \beta}_{\phantom{\alpha \beta};\mu}R_{\alpha \beta;\nu}- \frac{14}{15} \, R^{\alpha \beta}_{\phantom{\alpha \beta};(\mu}R_{\nu)\alpha;\beta}  +\frac{221}{210} \,
R^\alpha_{\phantom{\alpha} \mu;\beta} R_{\alpha \nu}^{\phantom{\alpha \nu};\beta} +
 \frac{113}{210} \,R^\alpha_{\phantom{\alpha} \mu;\beta} R^\beta_{\phantom{\beta} \nu;\alpha} +\frac{5}{21}\,
R^{\alpha \beta;\gamma} R_{\gamma \beta \alpha (\mu;\nu) } \right. \nonumber \\
&& \qquad \qquad \left.  -\frac{107}{210} \, R^{\alpha \beta;\gamma} R_{\alpha \mu \beta \nu;\gamma}-\frac{17}{210}\,R^{\alpha \beta \gamma \delta}_{\phantom{\alpha \beta \gamma \delta};\mu} R_{\alpha \beta \gamma \delta ; \nu }-\frac{29}{105} \, R^{\alpha \beta \gamma}_{\phantom{\alpha \beta \gamma}\mu;\delta}R_{\alpha \beta \gamma \nu}^{\phantom{\alpha \beta \gamma
\nu};\delta}  +\frac{5}{24}  \, R^2 R_{\mu\nu}\right. \nonumber \\
&& \qquad \qquad \left. -\frac{2}{5}\, R R_{\alpha \mu} R^\alpha_{\phantom{\alpha}\nu}
-\frac{31}{60} \, R^{\alpha \beta}R_{\alpha \beta}R_{\mu\nu} +\frac{1}{21} \, R^{\alpha \beta}R_{\alpha \mu}R_{\beta \nu}+ \frac{33}{35} \,R^{\alpha \gamma}R^\beta_{\phantom{\beta} r}R_{\alpha \mu \beta \nu}\right. \nonumber \\
&& \qquad \qquad \left.  - \frac{139}{105} \, R^{\alpha \beta}R^\gamma_{\phantom{\gamma} (\mu}R_{ |\gamma \beta \alpha|  \nu) }
- \frac{19}{30} \, R R^{\alpha \beta \gamma}_{\phantom{\alpha \beta \gamma}\mu }R_{\alpha \beta \gamma \nu} + \frac{1}{10}\,R_{\mu\nu}R^{\alpha \beta \gamma \delta} R_{\alpha \beta \gamma \delta }\right. \nonumber \\
&& \qquad \qquad\left. -\frac{74}{105}\, R^\alpha_{\phantom{\alpha}  (\mu}R^{\beta \gamma \delta}_{\phantom{\beta \gamma \delta} |\alpha|
}R_{ |\beta \gamma \delta| \nu) } -\frac{5}{42} \, R^{\alpha \beta}R^{\gamma \delta}_{\phantom{\gamma \delta} p\mu}R_{\gamma \delta \beta \nu}+ \frac{74}{105} \, R_{\alpha \beta}R^{\alpha \gamma \beta \delta}R_{\gamma \mu \delta \nu}\right. \nonumber \\
&& \qquad \qquad \left. -\frac{71}{105} \, R_{\alpha \beta}R^{\alpha \gamma \delta}_{\phantom{\alpha \gamma \delta}\mu}R^\beta_{\phantom{\beta} \gamma \delta \nu}+ \frac{37}{105} \,R^{\alpha \beta \gamma \delta}R_{\alpha \beta \sigma \mu }R_{\gamma \delta \phantom{\sigma}
\nu}^{\phantom{\gamma \delta} \sigma}\right. \nonumber \\
&& \qquad \qquad \left.
 + \frac{97}{105} \, R^{\alpha \gamma \beta \delta}R^\sigma_{\phantom{\sigma}
\alpha \beta \mu}R_{\sigma \gamma \delta \nu}  -\frac{37}{105} \,  R^{\alpha \beta \gamma}_{\phantom{\alpha \beta \gamma} \delta } R_{\alpha \beta \gamma \sigma}R^{\delta
\phantom{\mu} \sigma}_{\phantom{\delta} \mu \phantom{\sigma} \nu} + \frac{1}{5} \, R R^{\alpha \beta}R_{\alpha \mu \beta \nu}\right. \nonumber \\
&&  \qquad \qquad \left.
+ g_{\mu\nu} \left[  \frac{9}{280} \, \Box \Box R - \frac{19}{120} \, R\Box R -\frac{1}{84} \, R_{;\alpha \beta}
R^{\alpha \beta} -\frac{223}{420} \,  R_{\alpha \beta} \Box R^{\alpha \beta}+ \frac{79}{105} \, R_{\alpha \beta ; \gamma \delta}R^{\alpha \gamma \beta \delta} \right. \right.\nonumber \\
&& \qquad \qquad \left. \left. +\frac{163}{680}  \,R_{;\alpha}R^{;\alpha}-\frac{17}{56} \, R_{\alpha \beta;\gamma} R^{\alpha \beta;\gamma} -\frac{11}{420} \, R_{\alpha \beta;\gamma}
R^{\alpha \gamma;\beta} + \frac{51}{560} \,R_{\alpha \beta \gamma \delta;\sigma} R^{\alpha \beta \gamma \delta;\sigma}
     -\frac{5}{144}  \, R^3  \right.\right. \nonumber \\
&& \qquad \qquad \left.\left. +\frac{31}{120}\, RR_{\alpha \beta} R^{\alpha \beta} + \frac{1}{630} \,R_{\alpha \beta}
R^{\alpha}_{\phantom{\alpha} r}R^{\beta \gamma}  - \frac{53}{105} \, R_{\alpha \beta}R_{\gamma \delta}R^{\alpha \gamma \beta \delta} -\frac{1}{20}\,
RR_{\alpha \beta \gamma \delta} R^{\alpha \beta \gamma \delta} \right. \right. \nonumber \\
&& \qquad  \qquad \left. \left.
 + \frac{2}{5}\,R_{\alpha \beta}R^\alpha_{\phantom{\alpha} \gamma \delta \sigma} R^{\beta \gamma \delta \sigma } -\frac{263}{2520}\, R_{\alpha \beta \gamma \delta}R^{\alpha \beta \sigma \rho}
R^{\gamma \delta}_{\phantom{\gamma \delta} \sigma \rho} -\frac{106}{315} \, R_{\alpha \gamma \beta \delta} R^{\alpha \phantom{\sigma}
\beta}_{\phantom{\alpha} \sigma \phantom{\beta} \rho} R^{\gamma \sigma \delta \rho}\right ]\right].
\end{eqnarray}

The above result is valid for any static, stationary or general time-dependent spacetimes \cite{belokogne}. As we can see, it consist in a rather local complex expression for the renormalized stress energy tensor for a massive vector field in the Schwinger-DeWitt approximation. All the information coming from the massive vector field is included in the coefficients accompanying each local geometric term constructed from the Riemmann tensor, its covariant derivatives and contractions. As the only parameter that describes the geometry of a pointlike global monopole background is that determining the angle deficit, i.e, $\alpha$, we can expect that the renormalized stress tensor for the massive field in this spacetime be a function of $r$ and the parameters $\alpha$ and $m$.

Using (\ref{metric}) in (\ref{SET}) we obtain, for the components of the renormalized stress energy tensor for the massive spinor field in the pointlike monopole spacetime the very simple result
\begin{equation}
\left<T_{t}^{t}\right>_{ren}=\frac{\left(1-\alpha ^2\right)}{3360 \pi ^2 m ^2 r^6} \left[25 \alpha ^2\left(\alpha ^2+1\right)+4\right]
\label{eden}
\end{equation}
\begin{equation}
\left<T_{r}^{r}\right>_{ren}=\left<T_{t}^{t}\right>_{ren}
\label{p1}
\end{equation}
and
\begin{equation}
\left<T_{\theta}^{\theta}\right>_{ren}=\left<T_{\varphi}^{\varphi}\right>_{ren}=-2\left<T_{t}^{t}\right>_{ren}=\frac{\left(1-\alpha ^2\right)}{1680 \pi ^2 m ^2 r^6} \left[25 \alpha ^2\left(\alpha ^2+1\right)+4\right]
\label{p2}
\end{equation}

\begin{figure}[t]
\scalebox{0.57}{\includegraphics{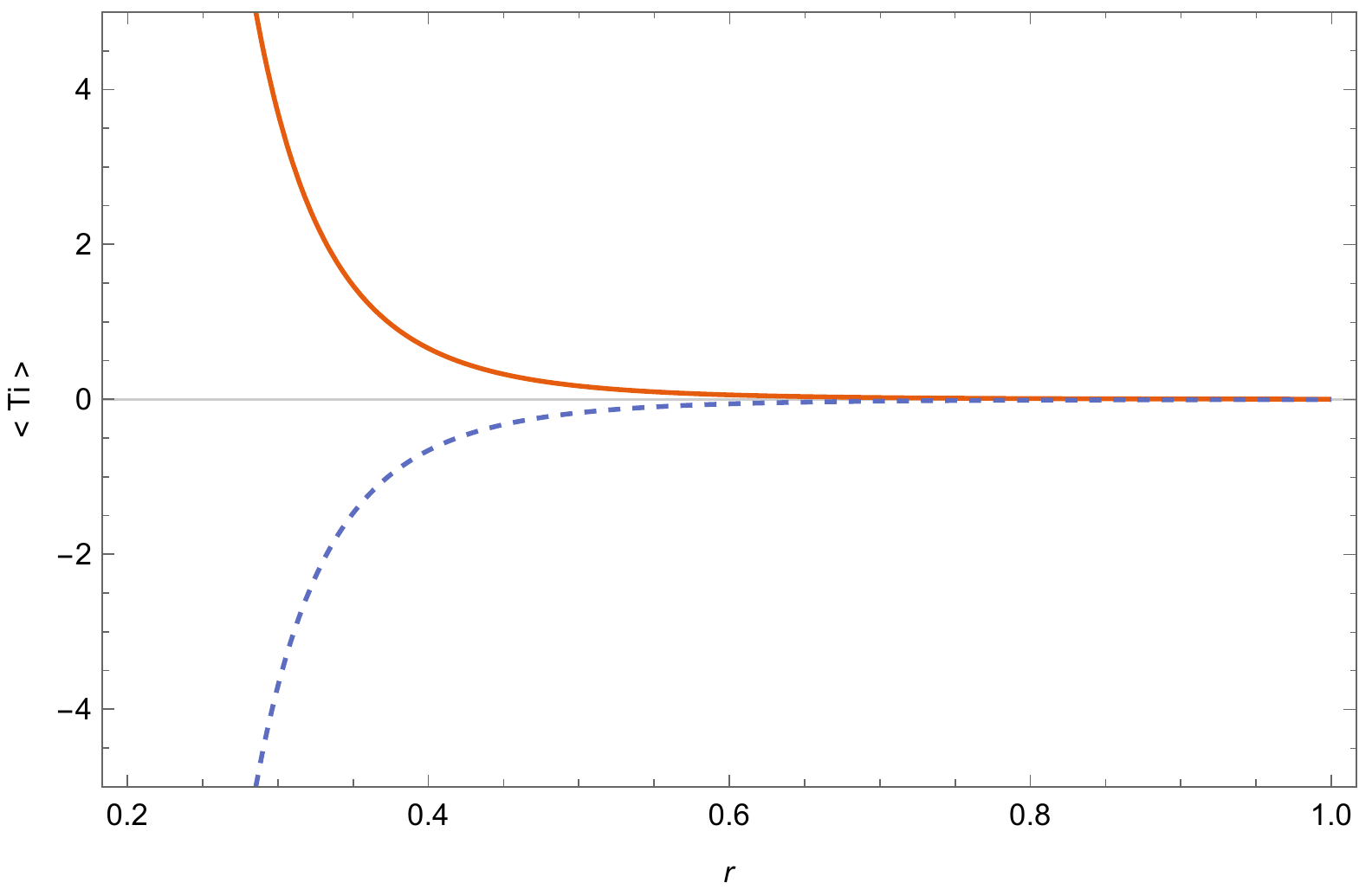}}
\caption{\textit{Dependance on the distance from monopole's core $r$ of the re-scaled time and radial components $\left<T_{0}\right>=3360\pi^{2}m^{2}\left <T_{t}^{t}\right >_{ren}=3360\pi^{2}m^{2}\left <T_{r}^{r}\right >_{ren}$ (solid line), and angular components $\left<T_{2}\right>=1680\pi^{2}m^{2}\left <T_{\theta}^{\theta}\right >_{ren}=1680\pi^{2}m^{2}\left <T_{\varphi}^{\varphi}\right >_{ren}$ (dashed line),of the renormalized stress energy tensor for a massive vector field in the pointlike global monopole spacetime. The values of the parameter used in the calculations is $1 - \alpha^2 =10^{-5}$ }.} \label{f1}
\end{figure}

In Figure (\ref{f1}) we show the dependance on the distance from monopole's core $r$ of the re-scaled time and radial components $\left<T_{0}\right>=3360\pi^{2}m^{2}\left <T_{t}^{t}\right >_{ren}=3360\pi^{2}m^{2}\left <T_{r}^{r}\right >_{ren}$, and angular components $\left<T_{1}\right>=1680\pi^{2}m^{2}\left <T_{\theta}^{\theta}\right >_{ren}=1680\pi^{2}m^{2}\left <T_{\varphi}^{\varphi}\right >_{ren}$, for a massive vector field in the pointlike global monopole spacetime.

As we see from the figure, the re-scaled time and radial components $\left<T_{0}\right>$ are decreasing functions of the distance from monopole's center, tending to zero as $r\rightarrow \infty$. Thus, the time and radial components of the renormalized stress energy tensor for the massive vector field in the pointlike global monopole background are positive at any distance $r$ from monopole's center.

An opposite behaviour present the re-scaled angular components $\left<T_{2}\right>$, which increases with negative values as $r$ increases, until it reach its maximum value equal to zero at large distances.  Thus, the angular components of the renormalized stress energy tensor for the massive vector field in the pointlike global monopole background are negative for all values of the distance $r$ from monopole's core.

The above results can be used to investigate the fulfillment of the energy conditions for the massive vector field in the pointlike global monopole spacetime. This energy conditions are restrictions need to be satisfied by the components of the stress energy tensor of matter fields, and are of two types, the pointwise energy conditions and the average energy conditions \cite{visser,matyjasekEC}. In the following we will analyse only the first type, that are of local character. While the fulfillment of the weak, strong, null and dominant energy conditions at the classical level are perfectly reasonable assumptions, semiclassical effects are capable of violating all of it.

Both local and averaged energy conditions are important concerning classical singularity theorems as the Penrose and the Hawking-Penrose singularity theorems, whose proof invokes the weak and strong energy conditions, respectively. Also the proof of the zeroth law of black hole thermodynamics relies on the dominant energy condition whereas in the proof of the second law of black hole thermodynamics one uses the null energy condition.

As usual we can define the energy density of the quantum massive vector field as $\rho=-\left<T_{t}^{t}\right>_{ren}$. The radial pressure is defined as $p_{r}=-\left<T_{r}^{r}\right>_{ren}$ and angular pressures as $p_{\theta}=p_{\phi}=p=\left<T_{\theta}^{\theta}\right>_{ren}=\left<T_{\varphi}^{\varphi}\right>_{ren}$.

The restrictions on the stress energy tensor for the quantum field that gives rise to the four pointwise energy conditions can be written in terms of the energy density and principal pressures defined above. The Null energy condition imply that $\rho-p_{r}\geq 0$ and $\rho+p\geq 0$. The Weak energy condition is equivalent to the Null energy condition with the extra constraint $\rho\geq 0$. The Strong energy condition is also equivalent to null energy condition supplemented with the constraint $\rho-p_{r}+2p\geq 0$, and finally the dominant energy condition is fulfilled when $\rho\geq 0$ and $-\rho\leq p_{i}\leq \rho$ \cite{visser}.

The results obtained in this work for the components of the renormalized stress energy tensor for a massive vector field in the spacetime of a pointlike global monopole imply that the energy density and the principal pressures of the quantum field are related by $\rho=p_{r}$ and $p=2\rho=2p_{r}$. Then, our results imply that, at all distances from the monopole's core, we have that $\rho\leq0$, $p_{r}\leq0$, $\rho-p_{r}=0$, $\rho+p=3\rho\leq0$ and $\rho-p_{r}+2p=4\rho\leq0$. Then, according to this relations, the quantum massive vector field in the pointlike global monopole backgreound violates all the pointwise energy conditions.

Another problem in which the results obtained in this paper can be used is the investigations of the changes in the geometry of the spacetime describing the pointlike global monopole induced by the non zero values of the components of the renormalized stress energy tensor of the massive vector field, the so called backreaction problem \cite{york,lousto-sanchez}. We can use $\left<T_{\mu}^{\nu}\right>_{ren}$ as a source in the semiclassical Einstei's equations, and solve it perturbatively. we can expect that as a result of the backreaction of the massive field upon the spacetime, the angle deficit characterizing this background changes. In a future report wi will address this problem.

\section*{Acknowledgments}
This work has been supported by TWAS-CONACYT $2017$ fellowship, that allow to the author to do a sabatical leave at Departamento de F\'isica Te\'orica, Divisi\'on de Ciencias e Ingenier\'ias, Universidad de Guanajuato, Campus Le\'on. The author also express his gratitude to Professor Oscar Loaiza Brito, for the support during the research stay at his group, where this work was completed.


\end{document}